\documentclass[12pt]{article}
\usepackage{amsfonts,amsthm,amsmath,amssymb,mathrsfs,url}

\title{On the Question Why There Exists Something Rather Than Nothing}
\author{
Roderich Tumulka\footnote{Fachbereich Mathematik, Eberhard-Karls-Universit\"at, Auf der Morgenstelle 10, 72076 T\"ubingen, Germany. E-mail: roderich.tumulka@uni-tuebingen.de}
}
\date{January 5, 2015}

\newcommand{\M}{\mathscr{M}}

\begin{document}
\maketitle
\begin{abstract}
In my opinion, nothing useful has ever been written on the question in the title, and small is the contribution that I have to offer. I outline an explanation for why there is something rather than nothing, an explanation which, however, I believe is incorrect because it makes a certain empirical prediction (absence of qualia) that is incorrect. Nevertheless, it may be interesting to discuss this reasoning. It allows, in principle though not in practice, to derive the laws of nature and all physical facts about the universe. Then I elucidate which objections to this explanation are, in my opinion, valid and which are not.
\end{abstract}

\section{Introduction}

Explanation in physics usually works this way: observable phenomena get explained by physical theories. A physical theory is the hypothesis that the physical world consists of certain kinds of physical objects governed by certain laws. From this hypothesis we derive, or we make it plausible that it can be derived, that the phenomenon in question (typically) occurs; then we say that the theory explains the phenomenon. The physical theory does not explain why these kinds of physical objects exist, why others do not, and why these laws hold; instead of explaining them, the theory merely posits them. At best, some theories are simpler and more elegant than others (e.g., Einstein's general relativity more than Newton's theory of gravity). But no physical theory comes close to explaining why these laws hold,  or why there exist any physical objects at all. Thus, no physical theory contributes to the question in the title.

Specifically, some physical theories allow for the possibility of a vacuum state, i.e., that at a certain time there is no matter in space, and for the possibility of a transition from a vacuum state to a non-vacuum state, i.e., that at some other time there is some matter in space. While such a theory has some explanatory value, it does not touch upon the question in the title, as it does not explain the physical laws, nor why space-time exists,\footnote{I also call space-time, not only pieces of matter, a physical object.} nor why certain kinds of matter (described by certain kinds of mathematical variables) exist and others do not. For related discussion, see Holt's overview \cite{Holt} and Albert's critique \cite{Al12} of Krauss's book \cite{Kr12}.

In this paper I outline a novel type of explanation of why anything exists. Obviously, the reasoning is very different from usual physical theories. The explanation also aims to explain why the world is the way it is. In particular, it allows in principle to derive the laws of physics and other testable consequences. I will answer to objections that I think are not valid, and I will argue that the explanation I describe, although it is a coherent reasoning, ultimately fails for a reason connected to the mind--body problem (see \cite{Ch} and references therein): it predicts the non-existence of qualia.

Even if the explanation fails it may be worth considering because there are, as far as I am aware, no other explanations for why anything exists. Perhaps, the closest that any reasoning before came to such an explanation was Anselm of Canterbury's (1033--1109) ontological argument for the existence of god \cite{Ans}, endorsed by Leibniz (see \cite{Loe} for a critique). I believe that Anselm's argument is not coherent and mine is; but I will not discuss Anselm's argument here.

I have the sense that many scientists and philosophers think that the question in the title is meaningless and cannot be answered. On the contrary, I think that the question is meaningful, as I understand its meaning, and the explanation I describe suggests to me that a rational answer might be conceivable.

\section{The Explanation}
\label{sec:exp}

The reasoning begins with the facts of mathematics. There is no mystery about why they are true. Their truth lies in the nature of mathematics and is explained by their content. Even if the physical world had been created by a god, the facts of mathematics would have been facts prior to that act, that is, it would not have been within god's power to change these facts. Along with the mathematical facts come mathematical objects, such as the empty set, the set containing only the empty set, the ordinal numbers (an ordinal number is a set $\alpha$ of sets that is well-ordered by the $\in$ relation and $\in$-transitive, i.e., contains each of its elements as a subset), the natural numbers regarded as the finite ordinal numbers, and the things that can be constructed from the natural numbers. I take it that the mathematical objects exist in some sense. Not in the same way as physical objects, but in the way appropriate for mathematical objects.

Now suppose, for the sake of the reasoning, that the mathematical objects could naturally be arranged in a way that looks like a discrete space-time, say a discrete version of a four-dimensional manifold. For example, it can be argued that all mathematical objects are ultimately sets; and all sets naturally form a directed graph with the $\in$ relation as the edges; and all sets naturally form a hierarchy according to their type.\footnote{The type of a set \cite{Rus} is an ordinal number such that a set of type $\alpha$ contains only sets of type smaller than $\alpha$.} A graph is somewhat similar to a discrete space-time; indeed some proposed definitions of discrete space-time say that such a thing is a partially ordered set---which is more or less a directed graph. The type has certain similarities with time (see Remark~\ref{rem:analogies} in Section~\ref{sec:rem}). The example breaks down at some point, however, since I do not see any similarity between the sets of a given type and 3-dimensional space. But suppose that the mathematical objects did naturally form a discrete space-time $\M$.

Suppose further that there is a property $M$ that some mathematical objects have and others do not, and which bears a natural similarity to the property of being matter (I will come back to this in Remark~\ref{rem:analogies} below). Then some space-time points in $\M$ will be ``occupied by matter'' (in the sense of having the property $M$) and others will not. They may form world-lines or other subsets of $\M$. Thus, in a certain way of looking at the world of mathematics, it appears like a space-time with matter moving in there. For example, think of all mathematical objects as being sets, take $\M$ again to be the graph of all sets with the $\in$ relation, and let $M$ be the property of being an ordinal number. That is a simple property that some objects have and others do not. If we think of type as time then, since there is exactly one ordinal for every type, at every time there is exactly one space-time point containing matter---a particle world-line. The example is of limited use since it represents a space-time with a single particle. But suppose that for some natural property $M$, the pattern of matter in space-time is more complicated, more like patterns in the physical world.

Suppose further that in $\M$ and in this pattern of matter, there are some sub-patterns which, if they actually were the patterns of physical matter in a physical space-time, would constitute intelligent beings. Then I feel that we are justified to say that in some sense these beings actually exist, since they are made of the elements of $\M$, which do exist. Let me call these beings the mathians. To the mathians, $\M$ must appear like a physical space-time, as it is the space-time in which they live, and the objects with property $M$ must appear like matter, as it is the matter of which they are made.

If the mathians ask what the physical laws of their space-time are, their question can be answered by pure thinking, by pure mathematics. In fact, every historical fact of their universe is determined by pure mathematics, although the mathians may not be able to carry out the thinking needed to determine these facts. 

If the mathians ask why their universe exists, then their question can indeed be answered. The correct answer is that $\M$, and the pattern of matter in $\M$ and the mathians, exist because they \emph{are} mathematical objects, which exist by their nature. If the mathians ask why the matter in their universe is distributed in a certain way, then their question can be answered. The correct answer is that the distribution of matter in their universe follows logically from the mathematical facts, which are necessarily true. It seems plausible to me that the mathians may or may not be aware that they are the mathians, and that their universe is $\M$, and that there is a reason for why their universe exists and looks the way it does. Or some mathians may be aware and others may not. I see no reason why all mathians would have to be aware that they are mathians.

Now suppose that \emph{we} are mathians, and that our space-time is $\M$. Then there \emph{is} an explanation to why our universe exists, and to why matter is distributed in a certain way in it. That is the explanation I promised. Let me call it the $i$-explanation. The $i$-explanation is that the matter of which we consist ultimately consists of mathematical objects, and the facts of our universe are ultimately mathematical facts; and there is no mystery to why mathematical objects exist or to why mathematical facts are true.

\section{Remarks}
\label{sec:rem}

\begin{enumerate}
\item The statement that the matter of which we consist ultimately consists of mathematical objects may sound similar to statements such as that (i)~a particle is a unitary representation of the Lorentz group, or that (ii)~reality according to quantum field theory consists of field operators, or that (iii)~reality according to quantum physics consists of pure information (``it from bit''). But it is not similar. 

Statement (i) really means, I take it, that the possible types of physical particles in our universe correspond to the unitary representations of the Lorentz group; not that an individual particle literally \emph{is} one of these representations. According to the $i$-explanation, in contrast, an individual particle (if our universe contains point particles) at a particular point in time actually \emph{is} a certain mathematical object. This object, by the way, is not as simple and beautiful as a unitary representation of the Lorentz group; rather, it must be an exorbitantly complicated mathematical object that is unlikely ever to be individually considered by a mathematician because there are so many other objects with similar properties. 

Let me turn to statement (ii). Its meaning is unclear to me, as I do not understand how operators can be a mathematical description of matter. I do understand, in contrast, how a subset of space-time can be a mathematical description of matter, and that is what the $i$-explanation provides. 

Statement (iii) means, I take it, that physical facts are not objective but exist only if observed by intelligent beings. According to the $i$-explanation, in contrast, there are objective physical facts, whether observed or not, namely that space-time is given by $\M$ and that matter is located at those space-time points with property $M$.

\item There are two ways in which mathematical objects and facts are relevant to the physical world according to the $i$-explanation. First, they may form part of the physical world, and second, they may apply to the physical world. For example, suppose $\M$ was the graph of all sets with the $\in$ relation. The number 5, understood as an ordinal number, is a set and therefore a space-time point in $\M$. However, the number 5 may also apply to other space-time regions, for example because there are 5 particles in a certain space-time region, i.e., there are 5 sets with property $M$ in a certain family of sets. We are familiar with the second role of mathematical objects but not with the first. 

\item Needless to say, the $i$-explanation does \emph{not} claim that every physical theory (say, Newtonian mechanics) is realized in some (part of an) physically existing universe, in contrast to an idea proposed by Tegmark \cite{Teg98,Teg07}. For example, any particular conceivable universe governed by Newtonian mechanics is a mathematical object, and thus corresponds to a space-time point rather than to the whole universe or a region therein; in particular, the universe $\M$ is not governed by Newtonian mechanics.

\item It is useful to distinguish between a specific $i$-explanation and the $i$-framework. To specify an $i$-explanation, it is necessary to specify how to arrange the totality of mathematical objects as a discrete space-time, and to specify the property $M$. Put differently, it is necessary to specify an \emph{isomorphism}, or \emph{identification}, $i$ between the mathematical world and the physical world. I have not specified $i$, so I have actually not provided an $i$-explanation. I have only outlined how such an explanation would work if $i$ could be provided. That is, I have described the \emph{$i$-framework}.

As long as no candidate for $i$ has been specified, it does not perhaps seem particularly likely that such an isomorphism $i$ exists. The failure of the example involving the graph of all sets with the $\in$ relation, and my failure to come up with a better example, make it seem even less likely. Yet, it seems conceivable that $i$ exists, and it seems that if $i$ exists then the $i$-explanation does explain why there exists something rather than nothing.

\item If $i$ exists and we know what it is then it is possible, at least in principle, to derive all laws of physics. It is also possible in principle, in this case, to determine all historical facts, past and future, unless there are limitations to the mathematical facts we can find out about. However, it is conceivable that the computational cost of finding out about interesting historical facts by analyzing the corresponding mathematical facts is prohibitively expensive, and in particular that finding out facts about our future, such as next week's stock prices, by means of a computation of the corresponding mathematical facts necessarily takes longer than those facts to occur (in the example, longer than one week).

\item It follows in particular that any specific candidate for $i$ can be, at least in principle, tested empirically, as it entails a particular distribution of matter in space-time that can be compared to the one in our universe. But even without such test results, some candidates for $i$ may be more persuasive, or more attractive, than others on purely theoretical grounds. For example, we may judge candidates for $i$ by their elegance and would require, in particular, that $M$ be a very simple property: If its definition were enormously complicated then that would cost the explanation much of its explanatory value. Furthermore, an $i$-explanation would seem more compelling if $M$ could be argued to be, not any old mathematical property, but a very special one, one that plays a crucial mathematical role. For example, the property of a set to be well-ordered by $\in$ and $\in$-transitive is a very special property, as testified by the fact that mathematician have a name (ordinal number) for such sets; basically, this property is special because well-orderings play a crucial mathematical role in set theory. 

\label{rem:analogies} Furthermore, an $i$-explanation would seem more compelling if analogies can be pointed out between $M$ and the physical property (of a space-time point) of being occupied with matter. In the same vein, an $i$-explanation would seem more compelling if analogies can be pointed out between $\M$ and space-time. For example, if $\M$ is the graph of all sets directed by the $\in$ relation, I can point out that the concept of \emph{type} of a set \cite{Rus} bears some analogies with the concept of \emph{time}: Both are linearly ordered, and, in fact, since the type governs the order in which sets must be defined so as to allow only well-defined objects as elements when forming a set, it seems natural to think of the type as a kind of ``logical time of set theory.''
\end{enumerate}

\section{Objections}
\label{sec:obj}

Objection: To the question why anything exists, most of the argument is irrelevant. Already in the first paragraph of Section~\ref{sec:exp}, it was claimed that mathematical objects exist. Therefore, something exists, and the remainder of the argument did not contribute to answering the question in the title. On the other hand, the first paragraph of  Section~\ref{sec:exp} did more or less take for granted that mathematical objects exist, so did not explain why they exist, and did not answer the question in the title.

\bigskip

\noindent Answer: There are two relevant senses of existence: mathematical and physical. The title refers to physical existence. The first paragraph of Section~\ref{sec:exp} refers to mathematical existence. Why something exists mathematically is easy to explain: Mathematical objects do, as soon as they are conceivable. Why something exists physically is hard to explain. In particular, physical objects do \emph{not} exist physically as soon as they are conceivable. The argument takes the mathematical existence of mathematical objects for granted and aims at explaining the physical existence of physical objects. That is why it is not over after the first paragraph.

\[***\]

\noindent Objection: No statement about physical existence can follow from statements exclusively concerned with mathematics. That is a matter of elementary logic, like ``ought'' cannot follow from ``is.''

\bigskip

\noindent Answer: That is true only as long as nothing is known about the meaning and the nature of physical existence. The hypothesis of our reasoning is that the physical existence of a space-time point $x$ ultimately means the mathematical existence of a certain mathematical object $O$ corresponding to $x$, and that the physical existence of matter at $x$ ultimately means that $O$ has the property $M$. If this is the case then statements about physical existence can clearly follow from mathematical statements. 

\[***\]

\noindent Objection: The $i$-explanation aims at explaining the existence of a physical world, of space-time, matter in motion, and all facts that supervene on matter in motion. But it does not lead to qualia (i.e., conscious experiences, such as experience of the color red \cite{Ch}). 
The mathians do not have qualia (i.e., they are not conscious, they do not experience the color red). If they claim they do then they are mistaken. I know that I have qualia, so I cannot be a mathian. 
If the $i$-explanation were correct then qualia would not exist. Thus, the $i$-framework makes a prediction, the absence of qualia, that can be regarded as an empirical prediction and that our findings disagree with.

\bigskip

\noindent Answer: I think that this objection is a good argument against the $i$-explanation. It may be tempting to hypothesize that qualia ultimately \emph{are} mathematical properties, but that does not fit with the nature of qualia: no mathematical structure would explain the way the color red looks (see, e.g., \cite{Ch}), so the nature of an experience of red cannot be mathematical. So I agree that the $i$-framework makes a prediction, the absence of qualia, that is empirically false. I conclude that the $i$-explanation, while it is a possible explanation of the existence of the physical world, is not the correct explanation.

\bigskip

\noindent\textit{Acknowledgments.} I thank Ovidiu Costin, Robert Geroch, and Barry Loewer for comments on an earlier version of this paper. I acknowledge funding from  the John Templeton Foundation (grant no.\ 37433).

\end{document}